\documentclass[12pt]{article}
\usepackage{amsmath}
\usepackage{color} \usepackage{amssymb} \usepackage{graphicx}

\usepackage{pifont}

\begin{document}

\begin{center}
{
\LARGE
PAMELA through a Magnetic Lens
\\[8mm]
}
	
J.~P.~Roberts\footnote{E-mail: \texttt{jonathan.roberts@nyu.edu}}
\\[5mm]
{\small\it
Center for Cosmology and Particle Physics, Department of Physics and Astronomy,\\ New York University, 4 Washington Place, New York}\\
[1mm]
\end{center}
\vspace*{0.75cm}

\begin{abstract}
The PAMELA satellite has observed an excess in the positron to electron ratio above theoretical predictions in the range 10 - 100 GeV that increases with energy. We propose that the excess is not due to a change in the local interstellar spectrum, but is due to heliospheric modulation. We motivate this from the known form of the heliospheric magnetic field and predict that the excess will disappear when we enter a period of solar maximum activity.
\end{abstract}

\setcounter{footnote}{0}

\section{Introduction}

The PAMELA satellite \cite{Adriani:2009p431} has observed a rise in the positron to electron fraction that increases with energy above 10 GeV. This has been interpreted as a rise in the positron fraction in the local interstellar spectrum of cosmic rays and has generated a great deal of discussion of primary positron sources, from decaying or annihilating dark matter to positron acceleration in pulsars.

We propose that the rise in the positron fraction is not due to a primary positron source and that the positron excess is not present in the interstellar spectrum, but is instead a result of the passage of cosmic rays through the heliospheric magnetic field.

There are two regimes for interstellar cosmic ray transport into the centre of the solar system. At high energies (above 10~TeV) cosmic rays free stream through the solar system with little effect from the solar system magnetic field. At low energies (below 1GeV) the cosmic rays cannot penetrate the centre of the solar system directly. During periods of solar minimum (such as during the period of the PAMELA measurements) the heliospheric magnetic field is ordered over large distances \cite{Zhou:2009p1228,UlyssesBook} in a Parker spiral \cite{Parker:1958p1139}. The large scale field and solar wind shield the centre of the solar system from the low energy interstellar cosmic rays, and the local flux of these cosmic rays is suppressed with respect to the interstellar flux \cite{UlyssesBook,Ndiitwani:2005p2956}. These low energy cosmic rays traverse the heliosphere by a complex process of diffusion, convection and drifts which delay their transition to the centre of the solar system.

In the energy range between these two regimes there is a transition from one mode of transport to the other and therefore there must be a transition from a locally suppressed flux at low energy to the unsuppressed interstellar flux at high energy. We will show that the structure of the magnetic field provides a window in the shielding that is charge asymmetric - allowing a fraction of the positively charged cosmic rays directly into the centre of the solar system but repelling negatively charged cosmic rays. We will also show that we expect the cosmic ray flux to be affected by the heliospheric magnetic field up to at least 500 GeV. The size of the window is energy dependent, first appearing for particles with energies around 10 GeV and growing in size for particles with higher energies. This means that, within these transition energies, a larger proportion of positively charged interstellar cosmic rays are able to reach the central solar system than negatively charged interstellar cosmic rays, and this asymmetry will grow with increasing energy. We show that the form of this charge asymmetry is consistent with the positron excess observed by the PAMELA satellite.

In our model the positron excess is caused by the large scale structure of the heliospheric magnetic field. This allows us to make clear predictions for the behaviour of the observed excess in relation to the solar system magnetic field. As we transition from the current solar minimum into a solar maximum and the magnetic field becomes chaotic and disordered, the charge asymmetry will disappear. This is happens every 11 years and the last solar maximum occurred around 2000, so we expect the current solar minimum to end very shortly. When the solar cycle settles down into the next solar minimum with the opposite orientation of the magnetic field, we should observe an increase in the proportion of negatively charged particles to positively charged particles. If this change is observed it will be a clear sign that the positron excess observed by the PAMELA satellite was a result of the heliospheric magnetic field configuration. AMS will be able to test this hypothesis within its first months of running.

We start by briefly describing the heliospheric magnetic field at solar minimum in section \ref{magneticField}. We review the current understanding of cosmic ray propagation in the heliosphere in section \ref{propagation}, particularly focusing on the role of convection in suppressing the local flux of cosmic rays with respect to the interstellar spectrum. In section \ref{lens} we show that the current sheet in the solar system midplane provides a means for negatively charged particles to enter the central solar system whilst presenting a barrier to positively charged particles. We first use a crude approximation of a flat current sheet and then consider the more realistic scenario of an oscillating current sheet. This allows us to present a model that fits the bulk properties of the positron excess observed by the PAMELA experiment. In section \ref{conclusions} we summarise our conclusions.

\section{The heliospheric magnetic field \label{magneticField}}

In 1958 E.~N.~Parker proposed that the solar system magnetic field takes the form of an Archimedes spiral \cite{Parker:1958p1139}. The sun is a rotating dipole and the magnetic field is carried out into the solar system by the supersonic solar wind. As the solar wind travels out from the sun its rotational velocity slows with respect to the plasma at the surface of the sun, causing it to be retarded and dragging the magnetic field into a spiral. This continues out to the termination shock at around 80~AU, where the solar wind becomes subsonic.

The heliospheric magnetic field is described in the Parker model by:
\begin{equation}
{\mathbf B}=B_0 \frac{r_0^2}{r^2}\left[\hat {\mathbf e}_r - \frac{\Omega_\odot (r-b) \sin\theta}{V}\hat{\mathbf e}_\phi\right],
\end{equation}
where ($r,\theta,\phi$) are heliocentric spherical coordinates, $B_0$ is the magnetic field at $r_0$, $r_0$ is the radius of the source surface, $V$ is the solar wind speed, $b$ is the radius at which the field is purely radial and $\Omega_\odot$ is the equatorial rotation rate of the sun, $2.7^{-6}rad/s$. At solar minimum $V = 400$km/s up to moderate latitudes (within $\pm60^\circ$ of the ecliptic plane). $V$ is larger near the poles (around 800 km/s).

The Parker model is very successful at modeling the magnetic field at moderate latitudes during times of solar minimum activity. More sophisticated models that perform better at high latitudes and times of greater solar activity have since been proposed. One widely used model is due to Fisk \cite{Fisk:1996p2820}:
\begin{equation}
{\mathbf B}=B_0 \frac{r_0^2}{r^2}\left[\hat{\mathbf e}_r - \frac{r \omega_\theta}{V}\hat{\mathbf e}_\theta - \frac{(\Omega_\odot-\omega_\phi)r\sin\theta}{V}\hat{\mathbf e}_\phi\right],
\end{equation}
where $\omega_\theta$ and $\omega_\phi$ are the differential rotation rates in the $\theta$ and $\phi$ direction respectively \cite{Burger:2005p1579}. This accounts for the differential rotation rate of the surface of the sun and gives a much more complex structure for the magnetic field.

In both cases the magnetic field is essentially a spiral. Close to the sunÕs surface the field lines are oriented radially. At large radii they are predominantly azimuthal. In the most recent solar minimum the radial and azimuthal components are roughly equal at 1~AU and the field is dominantly azimuthal beyond 10~AU. The winding of the magnetic field causes it to remain strong to large radii, with the strength of the azimuthal component falling off as $1/r$. The solar magnetic field is also oppositely aligned in the northern and southern hemisphere, with a sheet of zero magnetic field separating the two, known as the heliospheric current sheet (HCS).

The sunÕs magnetic poles are not aligned with its axis of rotation. This means that the HCS is tilted at the sun and that this tilt varies periodically as the sun rotates. These variations are carried out into the solar system by the solar wind and this causes the current sheet to be rippled. As the variations are carried out radially the latitudinal extent of the current sheet is described by a tilt angle $\theta_t$. At times of solar minimum, the tilt angle is small, often around $5-10^\circ$. At times of solar maximum the tilt angle grows to large values, before the solar polarity flips and the new polarity is carried out to the heliopause by the solar wind  \cite{Hoeksema:1995p3091}. The solar magnetic field flips on average every 11 years. The last reversal was in 2000 so the current solar minimum is coming to an end. Solar activity is currently increasing and will be followed by a flip in the magnetic field.

\begin{figure}[!t]
\begin{center}
\includegraphics{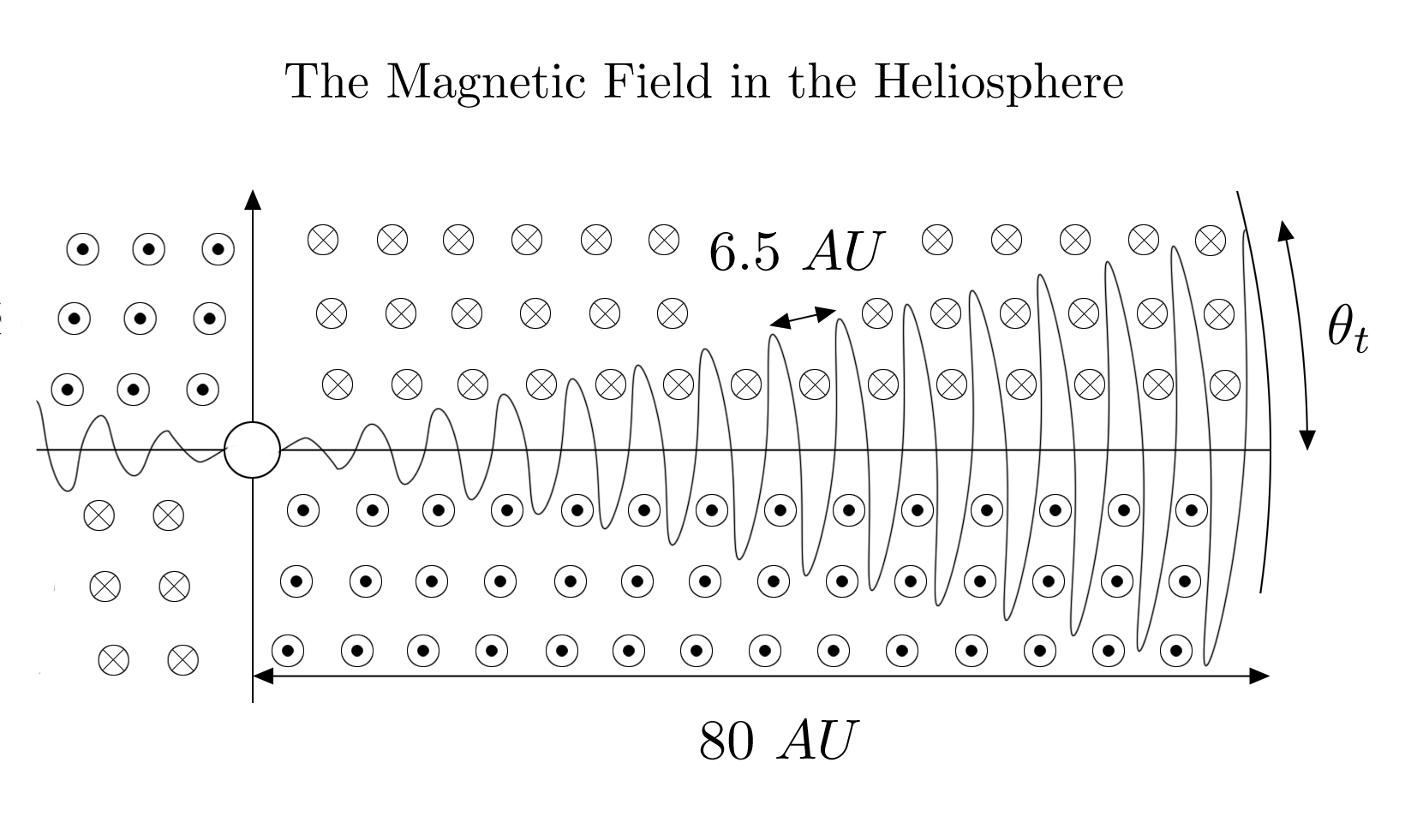}
\end{center}
\vspace*{-0.75cm}
\caption{
\small \it A schematic of the form of the heliospheric magnetic field in the current solar minimum. \label{MagneticField}}
\end{figure}

During the time of the the PAMELA measurements the heliosphere was in a period of solar minimum. The magnetic field configuration is shown in figure \ref{MagneticField} with the field lines directed towards the sun in the northern hemisphere and outward from the sun in the southern hemisphere. The local magnetic field at earth during the PAMELA measurements had a magnitude of around 3.3~nT and we take this value throughout for our model. For a good review of the heliospheric magnetic field see \cite{UlyssesBook}.

\section{Cosmic ray propagation through the heliosphere\label{propagation}}

To consider the motion of galactic cosmic rays\footnote{We consider only galactic cosmic rays here. Solar cosmic rays are much less energetic. Occasional solar flares can accelerate electrons to energies of a few hundred MeV, but even these are rare. Therefore in the energy range we consider (1 - 100 GeV) the electrons and positrons are solely galactic in origin.} through the heliosphere it will be useful to have the equation for the Larmor radius of a relativistic particle in a magnetic field in convenient units:
\begin{equation}
r_L(AU)=\frac{2.2\times 10^{-2} p_\perp (GeV/c)}{|q| B(nT)}.
\label{gyroradius}
\end{equation}

If we take the magnetic field strength (at low latitudes) to be $|B|=B_E/r(AU)$ and the local field strength at earth over the course of the PAMELA data to be $B_E=3.3~nT$ then we can approximate the Larmor radius of a particle at low latitudes to be:
\begin{equation}
r_L(AU)=6.67\times 10^{-3} p_\perp (GeV/c)r(AU),
\label{gyroradius2}
\end{equation}
where $r$ is the distance from the center of the solar system.

From this we can see that a singly charged particle with a rigidity of $1~GV$ has a Larmor radius of less that 1~AU even in the weak magnetic field near the termination shock at 80~AU. Such a particle cannot propagate directly to the centre of the solar system as $r_L < r$ at all points. At the other extreme, a 1~TeV particle has $r_L = 6.7r$ for all $r$. As $r_L \gg r$ at all points on its path, it can penetrate the centre of the solar system with relatively little deflection in its trajectory (around $10^\circ$). At energies above a few tens of TeV the deflection becomes negligible.

Between these two limits we must consider two separate populations of cosmic rays:
\begin{enumerate}
\item those that travel to the center of the solar system by traveling along magnetic field lines
\item those that can penetrate the center of the solar system by traveling perpendicular to the magnetic field.
\end{enumerate}

At high energies ($> 1$ TeV) the flux of cosmic rays at earth should correspond directly to the interstellar spectrum. The same cannot be said of the low energy cosmic rays. The transport of sub-GeV cosmic rays within the solar system is described by the transport equation Ñ first proposed by Parker in 1965 \cite{Parker:1965p2887}:
\begin{equation}
\frac{\partial U}{\partial t}=-\nabla\cdot(C{\mathbf V}U) - \nabla\cdot(\langle {\mathbf v}_d\rangle U) + 	\nabla\cdot({\mathbf \kappa}^{(S)}\cdot \nabla U) - \frac{1}{3}\frac{\partial}{\partial p}(p{\mathbf V}\cdot \nabla U),
\label{TransportEquation}
\end{equation}
where $U$ is the density of cosmic rays as a function of position $\mathbf{r}$, momentum $p$ and time $t$, ${\mathbf V}$ represents the solar wind velocity, $C=1-(1/3U)\partial/\partial p (pU)$ is the Compton-Getting coefficient, $\langle {\mathbf v}_d\rangle$ is the averaged drift velocity and ${\mathbf \kappa}^{(S)}$ is the symmetric part of the diffusion tensor:
\begin{equation}
{\mathbf \kappa}=  \begin{pmatrix}
    \kappa_{||} & 0 & 0 \\ 0 & \kappa_\perp & \kappa_T \\ 0 & -\kappa_T &
    \kappa_\perp
  \end{pmatrix}.
\end{equation}
Parallel $||$ and perpendicular $\perp$ are defined with respect to the magnetic field direction.

The terms on the right hand side of the transport equation represent the change in particle density over time. The first term represents convection caused by the solar wind carrying particles out of the solar system. This effect is large when the average radial velocity of a particle is of the same order or smaller than the solar wind speed. The second term quantifies the effects of drifts - where a variation in the magnetic field strength allows for a motion of the particles in the direction of $\nabla\times{\mathbf B}$. These drifts are large scale particle motions and are modeled as effective sources or sinks in the system. Drifts are charge asymmetric processes and are the cause of the suppression of the positron flux below 10~GeV. Drifts only affect the local particle density if $\nabla U$ is non-zero as they only appear in the transport equation as $\langle \mathbf{v}_d\rangle \nabla U$. At low energies convection is known to cause a radial gradient in the cosmic ray density with a local deficit, resulting in a large $\nabla U$. This determines the size of the effect of drifts on the local cosmic ray density. The third term covers diffusion - modeling the scattering of particles off inhomogeneities in the magnetic field. The terms in the diffusion matrix are not well known. For low energy transport, perpendicular diffusion is taken to be significantly less efficient than parallel diffusion ($\kappa_\perp\approx 0.02\kappa_{||}$). The final term represents adiabatic energy changes. The solution of this equation in 3D requires sophisticated numerical simulations which we will not discuss further here. For a review of recent work in this area  see \cite{Burger:2005p1579,Ferreira:2008p2404}.

For our purposes it is enough to note that the transport of low energy cosmic rays is very different from the transport of high energy cosmic rays. The average radial velocity of low energy cosmic rays is small - through perpendicular diffusion and drifts. As the magnitude of the inward radial velocity is smaller than the outward solar wind speed the convection effects of the solar wind dominate and create a radial gradient in the cosmic ray density. This is well measured for low energy cosmic rays ($E<200$~MeV)and was analysed over different solar minima and maxima in \cite{Fujii:2005p1432}. The gradient has also been measured for protons and electrons with higher rigidities along the Ulysses trajectory \cite{Ndiitwani:2005p2956} where they were shown to have a radial gradient of 2-3\%/AU at solar minimum. In contrast we know that there cannot be any radial gradient in the cosmic ray density at very high energies when the cosmic rays free stream through the solar system.

We take the local flux $F_{1AU}$ to be suppressed by an energy independent factor $A$ when compared to the interstellar flux $F_{80AU}$, up to energies at which particles can penetrate the center of the solar system directly (well above the highest energy PAMELA measurement):
\begin{equation}
F_{1AU}=AF_{80AU}\label{eq:suppression}
\end{equation}
This is an approximation as particles will penetrate further into the solar system as their energy increases. However for the energies we are interested in for PAMELA ($< 100$ GeV), we expect the approximation to be good for cosmic ray fluxes at earth because the Larmor radii are small compared to the distance travelled.

This energy dependent flux deficit would have little bearing on the PAMELA result if there was a smooth transition from one propagation method to the other. However, the form of the magnetic field allows some particles to penetrate directly into the central solar system even at low energies, and this effect is charge asymmetric.

\section{The magnetic lens \label{lens}}

\subsection{A flat current sheet\label{lens1}}

\begin{figure}[!t]
\begin{center}
\includegraphics{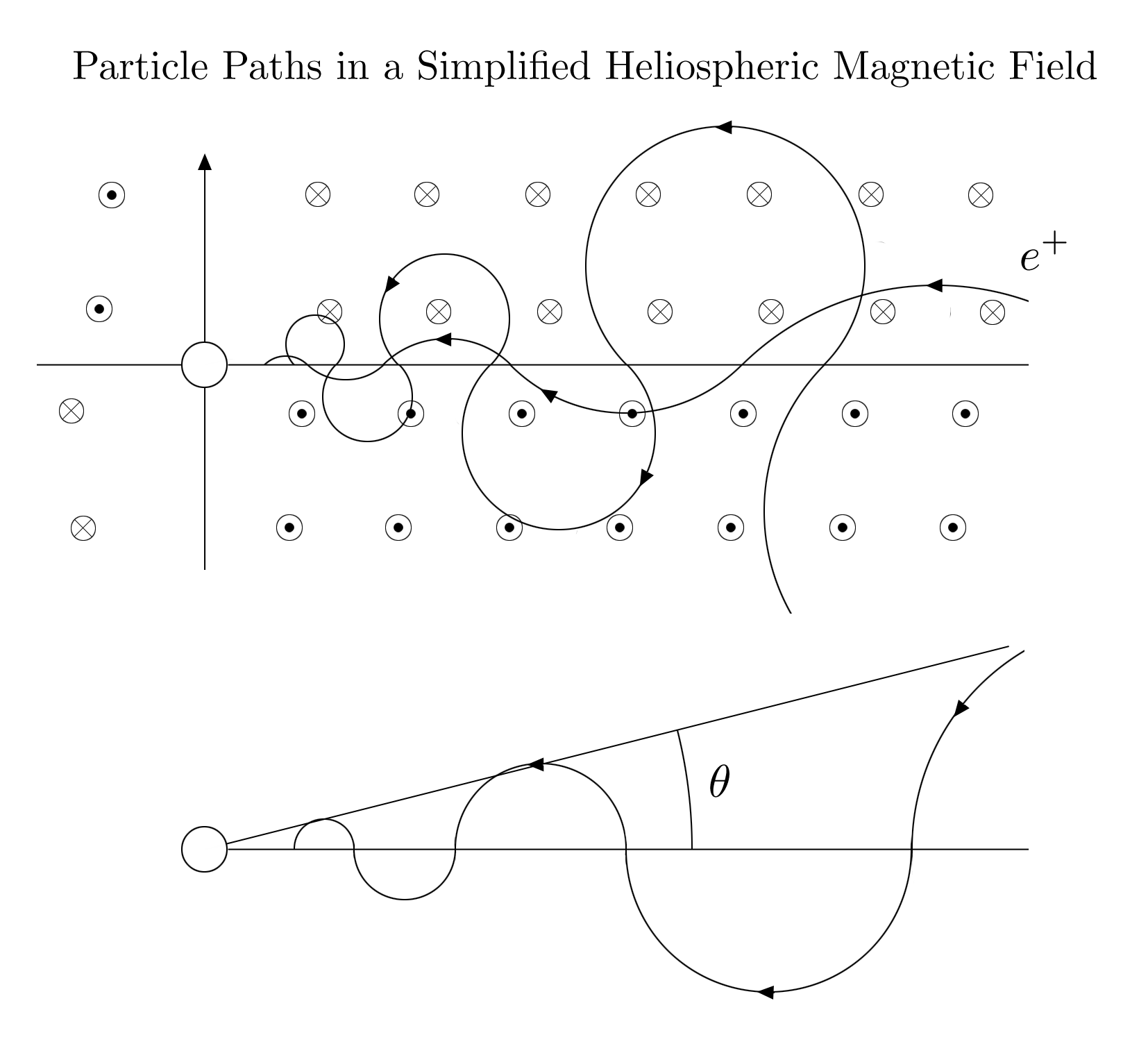}
\end{center}\vspace*{-0.75cm}
\caption{
\small \it Paths of positively charged particles that cross the ecliptic plane in the approximation of a flat current sheet and a simplified magnetic field. Note that it does not matter at what angle the particle crosses the ecliptic plane, the overall motion of a positively charged particle is always towards the centre of the solar system. Negatively charged particles have the opposite paths, drifting out of the solar system. The sun is to the left. The horizontal line is both the ecliptic plane and the heliospheric current sheet (HCS) in this simple model. \label{ParticlePaths1}}
\end{figure}

Let us consider a magnetic field that is purely azimuthal and has a strength that falls off as $B_E/r(AU)$. Let us further take the approximation of a flat current sheet in the ecliptic plane. In this magnetic field, the Larmor radius of the particle is given by:
\begin{equation}
r_L(AU)=\frac{2.2\times 10^{-2} p_\perp(GeV/c)r(AU)}{|q|B_E(nT)}.
\end{equation}

In figure~\ref{ParticlePaths1} we show a schematic of the paths of positively charged particles that cross the ecliptic in this toy model of the current magnetic field. Any positively charged particle that crosses the ecliptic will have an overall motion towards the centre of the solar system. For particles of the opposite sign the motion of a particle that crosses the ecliptic is away from the center of the solar system\footnote{We are not the first to notice this effect. It was first shown to exist by Levy in 1976 \cite{Levy:1975p2266,Levy:5p2267}. He noted that there should be a large charge asymmetry and that this should exist up to large rigidities - in his case he took it to be up to 30GV, with the high energy cut-off being due to a reduced radial gradients in cosmic ray density at these energies. Following LevyÕs work it was argued that the oscillation of the current sheet reduced the effectiveness of the drift along the current sheet, resulting in no modulation above a few GeV. We will address this point in section~\ref{lens2}.}. Particles that do not cross the ecliptic will complete a full Larmor rotation in the outer heliosphere. These particles must reach the centre of the solar system by the processes of convection and diffusion discussed in section \ref{propagation}. Therefore we can consider these to be two separate populations of cosmic rays. The one population of positively charged cosmic rays that takes a path such as that in figure \ref{ParticlePaths1}, and the second population that reaches the center of the solar system by the much longer (charge symmetric) path of diffusion, convection and drifts in which the local flux is suppressed by the convection effects discussed previously.

The fraction of the interstellar cosmic rays incident on the heliosphere that can follow the paths shown in figure \ref{ParticlePaths1} is proportional to momentum. As the Larmor radius increases with the distance from the centre of the solar system (due to the falling magnetic field) we can define the fraction of particles that can travel directly to the centre of the solar system by using the angle $\theta$ shown in figure \ref{ParticlePaths1}:
\begin{equation}
f(p)=\sin\theta=\frac{r_L(p)}{r}=\frac{2.2\times 10^{-2} p_\perp(GeV/c)}{|q|B_E(nT)}. \label{fp}
\end{equation}
$f(p)$ gives the fraction of the incident flux of positively charged interstellar cosmic rays that cross the current sheet and propagate directly to the centre of the solar system.\footnote{As we saw in figure~\ref{ParticlePaths1}, particles that are up to $2r_L$ from the ecliptic can cross the ecliptic and drift towards the center of the solar system. However as the distance from the current sheet increases, the range of angles that allow the particle to reach the current sheet decreases. The fraction of particles at a height $d$ from the ecliptic that have trajectories that can cross it is given by $1/\pi\cos^{-1}(d-r_L)$. When we integrate over this distribution we find that $1/2$ of all particles in the volume defined by the angle $-2\theta\rightarrow2\theta$ (see figure \ref{ParticlePaths1}) have trajectories that can cross the plane. This gives the total fraction of the particles to be $f(p)$ in eq.~(\ref{fp}).} As the momentum of the particle increases, the Larmor radius increases and the fraction of particles that can cross the heliospheric current sheet (HCS) increases. Thus the fraction is directly proportional to momentum.

This formula is not valid at large angles as we would need to carefully treat the varying Larmor radius of the particle over the course of one rotation due to the latitudinal variation in the magnetic field. Here we are considering only particles with energies within the range of the PAMELA measurements. The upper reach of PAMELAÕs positron fraction data of 100 GeV corresponds to an angle from eq.~(\ref{fp}) of 33 degrees, small enough that these corrections can be neglected in our first approximation.

The fraction of the interstellar spectrum that can propagate directly to the centre of the heliosphere through this magnetic lens provides an additional component to the local cosmic ray flux. Therefore we add an additional term to eq. (\ref{eq:suppression}) for the positron flux:\footnote{Here we consider electrons and positrons, but we note that the same arguments must apply to other cosmic ray species. Specifically we would expect to see a deficit of anti-protons in the current solar minimum. For protons and anti-protons we cannot approximate rigidity with energy and we expect the rise to be slower and extend to lower energies. We will present a careful analysis of the proton to anti-proton flux in a future paper but for now we note that this model requires that there be no anti-proton excess in contrast to many dark matter explanations of PAMELA.}
\begin{eqnarray}
\nonumber F_{1AU}(e^+)&=&(A+f(p))F_{80AU}(e^+)\\
F_{1AU}(e^-)&=&A F_{80AU}(e^-)
\label{FlatFlux}
\end{eqnarray}

\begin{figure}[!t]
\begin{center}
\includegraphics{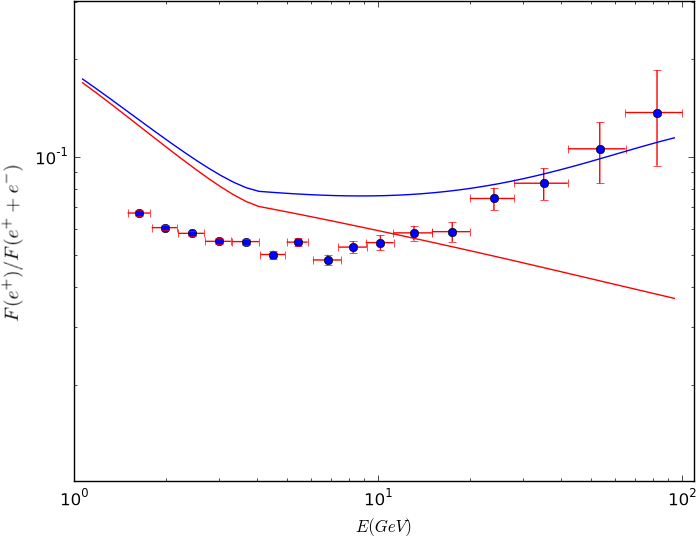}
\end{center}
\caption{\small \it  We use a standard background spectra generated using Galprop \cite{Strong:1998p3045,Cholis:2009p3109} - plotted in red (dashed). The resulting positron excess, taking into account the effect of the current sheet from the simple model of eq. (\ref{FlatFlux}), is shown in blue (solid). We compare this with the spectrum from the PAMELA experiment (blue circles with error bars). \label{PAMELAFit1}}
\end{figure}

Using this model we compare a modified spectrum to the PAMELA data in figure \ref{PAMELAFit1} with $A=0.23$. The general behaviour is consistent with the rising profile of the high energy fraction but it does not correctly model the positron fraction below 30~GeV, instead giving a positron fraction that is significantly too high.

\subsection{An oscillating current sheet \label{lens2}}

In the previous section we considered a simplified magnetic field with a flat current sheet. This is not realistic - we know that the current sheet oscillates and extends to large heliolatitudes. Even at solar minimum the current sheet extends up to around $15^\circ$ (from the Wilcox Solar Observatory \cite{WSO}). The tilt varies considerably from month to month. However we will consider an idealised situation where the tilt remains constant throughout the heliosphere to qualitatively understand the effect of the tilted current sheet on the high energy cosmic ray flux.

\begin{figure}[!t]
\begin{center}
\includegraphics{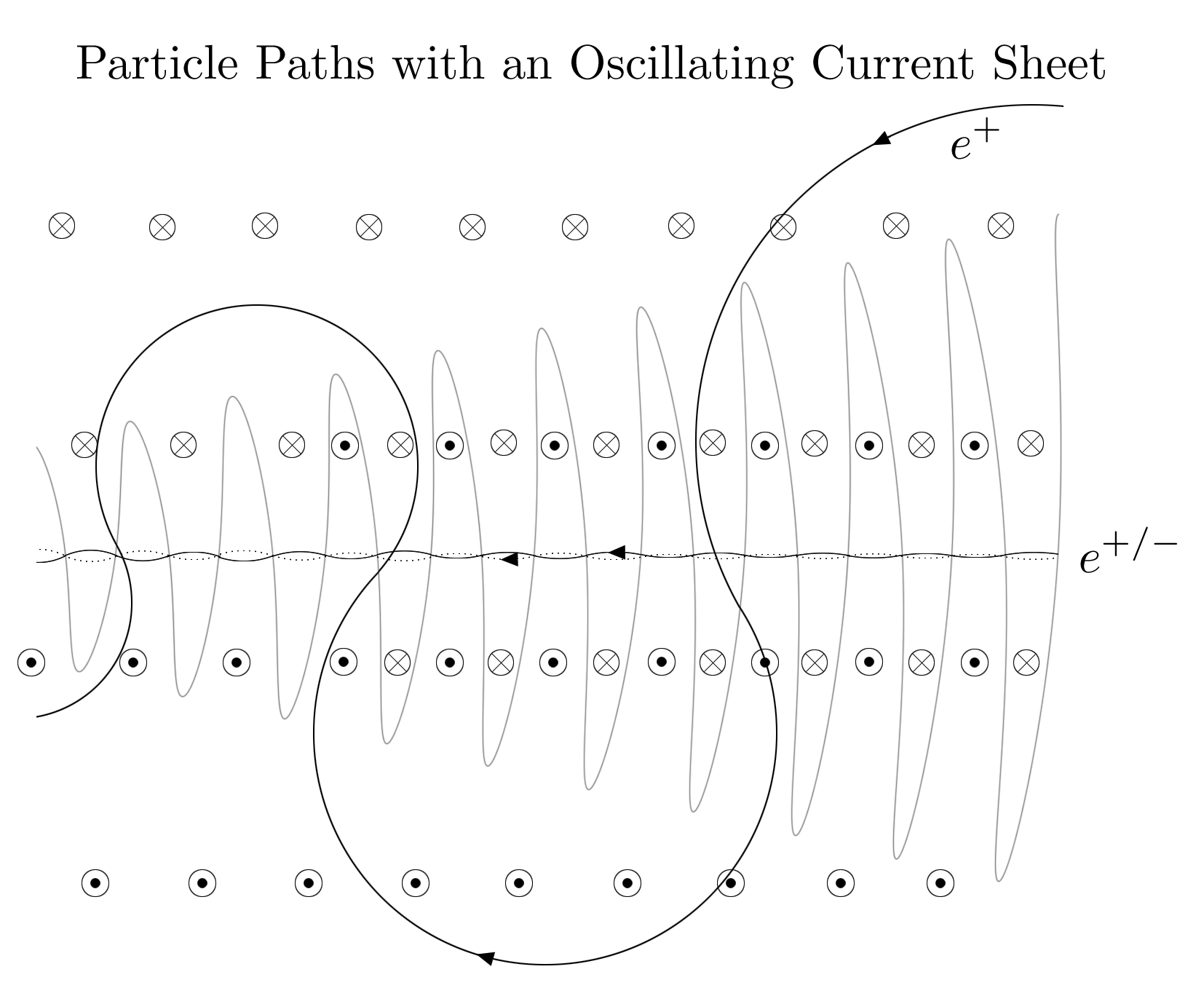}
\end{center}
\caption{\small \it A schematic diagram of the particle paths in the magnetic field with an oscillating current sheet. The light grey line with increasing amplitude depicts the current sheet. Solid black lines show paths of positively charged particles. The dotted line shows the path of a negatively charged particle. The centre of the heliosphere is to the left, so the magnetic field strength increases to the left. Particles of both signs can penetrate the centre of the solar system along the ecliptic plane once their momentum is large enough that their Larmor radius exceeds the wavelength of the current sheet. Particles with a large Larmor radius that leave the angular volume described by the current sheet tilt $\theta_t$ have paths that are equivalent to the scenario of a flat current sheet. \label{ParticlePaths2}}
\end{figure}

`Tilt' is a slightly misleading term. The `tilt' refers to the magnitude of the oscillations of the current sheet which are generated by the angle of the magnetic poles of the sun to its axis of rotation. This leads to a corrugated geometry of the current sheet shown in figure \ref{ParticlePaths2}.

In modulation theory \cite{Burger:1989p2324} the effect of the oscillating current sheet has been incorporated by first taking it to be a flat sheet that gives rise to an overall drift velocity in the ecliptic plane. We saw this overall velocity in the toy model of figure 2 with the average velocity of a positively charged particle being along the flat current sheet towards the centre of the heliosphere. In the case of a corrugated current sheet the average drift velocity is still along the current sheet Ñ but a particle following the current sheet now has to travel much further to get to the centre of the heliosphere. This means the average radial velocity is lower than the solar wind speed and the inward drift of the cosmic rays is countered by the outward convection of the solar wind.

The approximation that particles follow the current sheet works well as long as $r_L$ is smaller than the wavelength of the oscillations of the current sheet (around 6.5~AU). At higher energies this is no longer true. We show the particle paths of higher energy particles in figure \ref{ParticlePaths2}. When $r_L$ exceeds the wavelength of the oscillations of the current sheet a particle can pass directly along the ecliptic Ñ as shown in figure \ref{ParticlePaths2}. This is charge {\it symmetric} close to the ecliptic. This allows particles of both species to propagate deep into the heliosphere. As the Larmor radius goes as $r$, and the wavelength of the HCS remains constant with $r$, a particle with a given energy will penetrate up to the point where $r_L < 6.5$~AU. At this point a positively charged particle will continue to travel inwards along the current sheet, and a negatively charged particle will not.

High energy particles that travel outside the volume containing the current sheet still have the same asymmetric paths that we saw in section \ref{lens1} that cause positrons to move to the centre of the solar system and cause electrons to move out. This is to be expected, as when $r_L$ is larger than the amplitude of oscillation there are many paths along which the positrons cross the current sheet only once per rotation. For these particles the oscillation of the current sheet is irrelevant and the effect is identical to the case of a flat current sheet considered earlier.

To accurately account for the variation in the cosmic ray species due to these complex paths will require a detailed computer model of the particle propagation. This is underway and will be presented in a future work. Here we construct an approximation to highlight the main effects of the oscillating current sheet. The current sheet defines two volumes in the heliosphere:
\begin{itemize}
\item The volume containing the current sheet, defined by the angle $\theta_t$ in figure \ref{ParticlePaths2}. Within this volume, particles of both charges can propagate to the center of the solar system as long as their Larmor radius is greater than the wavelength of the current sheet, $r_L > 6.5$~AU . Particles that remain within this volume do not contribute to any charge asymmetry.
\item The volume defined by the Larmor radius of the propagating particle, delimited by the angle $\theta(p)$ from figure \ref{ParticlePaths1}. If $\theta < \theta_t$ then the particles remain within the volume defined by the current sheet and donÕt contribute to any charge asymmetric flux. If $\theta > \theta_t$ particles can propagate outside the volume defined by the current sheet. Those that travel outside the volume cross the current sheet only once per half rotation - as shown in figure \ref{ParticlePaths2} - and provide a charge asymmetry as in the case of a flat current sheet.
\end{itemize}

Therefore the oscillating current sheet has the effect of reducing the fraction of particles that are lensed by the magnetic field, when compared to the flat current sheet considered in section \ref{lens1}. The new fraction $f'(p)$ is given by:
\begin{equation}
f'(p)=f(p)-\delta,~~~f'\geq 0
\end{equation}
where $\delta$ is proportional to $\sin\theta_t$, as the reduction is directly due to the latitudinal extent of the current sheet.

Taking this into account we now write the ratio of fluxes to be:
\begin{equation}
\frac{F_{1AU}(e^+)}{F_{1AU}(e^++e^-)}=\frac{(A+f'(p))F_{80AU}(e^+)}{((A+f'(p))F_{80AU}(e^+)+AF_{80AU}(e^-)}.
\label{WavyFlux}
\end{equation}

\begin{figure}[!t]
\begin{center}
\includegraphics{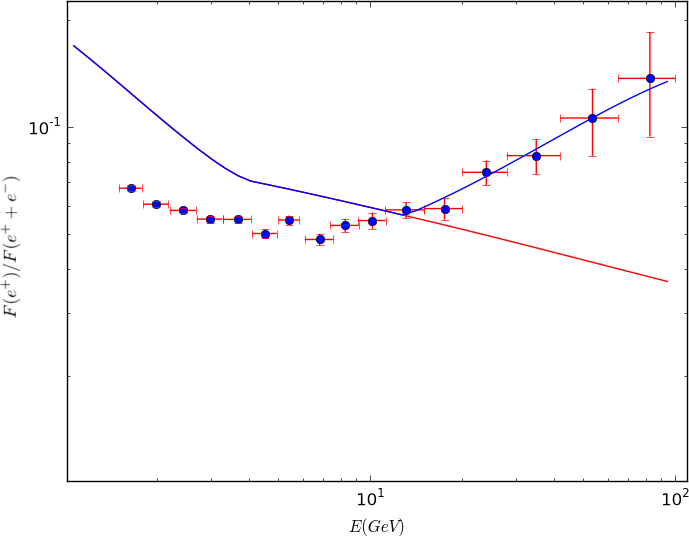}
\end{center}
\caption{\small \it The PAMELA data (blue circles with error bars) compared to a model that incorporates a varying current sheet (solid line) and the unmodified standard Galprop prediction (red dashed line). At low energies (below 10GeV) we do not expect a good agreement as conventional heliospheric modulation effects dominates at low energies.\label{PAMELAFit2}}
\end{figure}

We show the result in figure~\ref{PAMELAFit2}. This plot requires $A = 0.15$ and for $\delta = 0.09$. The oscillating current sheet has the effect of moving the rise in the positron fraction to larger energies, and decreasing the overall magnification effect of the lens.

Even though we have used a number of broad approximations, the predictions are robust. There is an asymmetry in the propagation paths of particles of different signs at the energies measured by PAMELA. This asymmetry only becomes significant once the Larmor radius of a particle is large with respect to the latitudinal extent of the current sheet, and once it becomes significant the effect increases with energy. We only need to assume that there remains a radial gradient in the density of cosmic rays at these energies\footnote{The radial gradient in cosmic ray densities should disappear for energies of a few TeV when particles free stream through the heliosphere. It is tempting to use this to explain the Fermi excess in cosmic ray electrons. In this case the bump at 500~GeV would be the observation of the unmodulated spectrum whereas the low energy data would represent the locally suppressed flux. To fit the Fermi data in this way would indicate the local flux at 10 GeV to be 60\% of the unmodulated interstellar flux - 4 times the flux we use to fit the PAMELA signal. We note that there are many sources of uncertainty in our model that could account for this discrepancy and we will require that our more complete description should account for this difference. There also remains the possibility that there is an extra primary component to the electron and positron spectrums but that such an excess would be smaller than previously considered.} to predict a rising positron fraction between 10 and 100~GeV.

\section{Conclusions \label{conclusions}}

We have proposed that the PAMELA positron excess is not down to annihilating dark matter or new astrophysical sources, but instead results from the configuration of the heliospheric magnetic field. The ordered nature of the magnetic field on large scales creates a lens that allows particles of one sign to propagate into the centre of the solar system whilst particles of the opposite sign travel out. This effect rises with energy and has the correct sign for the configuration of the magnetic field during PAMELAÕs data taking. It naturally occurs in the correct energy range to account for the observed positron excess.

Though our model of the heliospheric magnetic field is rough, we can already make three clear predictions. Firstly, we predict that an increase in the tilt angle of the solar magnetic field will result in an energy dependent decrease in the positron excess. Secondly, in a period of solar maximum the lack of any large scale ordered magnetic field will cause the positron excess to disappear. This suppression will be delayed by about a year as the increased activity at the sun propagates out to the edges of the heliosphere. Finally, in the following solar minimum with the opposite magnetic field orientation we predict an electron excess as the electrons are focused by the magnetic lens and the positrons are repelled. These signals will be tested by AMS within its first months of running.

\section*{Acknowledgements}
The work of JPR was supported by NSF Awards PHY-0758032, PHY-0449818 and NSF 0900631 and DoE Award No. DE-FG02-06ER41417. He would like to thank R. Johnson and E. J. Smith for advice on the heliospheric magnetic field, Ilias Cholis for discussions on interstellar spectra, Andre Gruzinov for discussions about LiouvilleÕs theorem and Neal Weiner for many useful discussions about the PAMELA data. He would also like to thank the GGI and the INFN for their hospitality in Florence when this work was being completed.

\bibliographystyle{unsrt}
\bibliography{references2}

\end{document}